\documentclass[aps,pra,reprint,superscriptaddress,groupedaddress]{revtex4-1}  % for review and submission

\usepackage{graphicx}  % needed for figures
\usepackage{dcolumn}   % needed for some tables
\usepackage{bm}        % for math
\usepackage{amssymb}   % for math
\usepackage{amsmath}
\usepackage{graphicx}
\usepackage{float}
\begin{document}

\title{Pseudospin Electronics in Phosphorene Nanoribbons}

\author{S. Soleimanikahnoj}\email{soleimanikah@wisc.edu}
\author{I. Knezevic}
\email{iknezevic@wisc.edu}
\affiliation{Department of Electrical and Computer Engineering, University of Wisconsin-Madison, Madison,
WI 53706, USA}

\date{\today}

\begin{abstract}

Zigzag phosphorene nanoribbons are metallic owing to the edge states, whose energies are inside the gap and far from the bulk bands. We show that -- through electrical manipulation of edge states -- electron propagation can be restricted to one of the ribbon edges or, in case of bilayer phosphorene nanoribbons, to one of the layers. This finding implies that edge and layer can be regarded as tunable equivalents of the spin-one-half degree of freedom i.e., the pseudospin. In both layer- and edge-pseudospin schemes, we propose and characterize a pseudospin field-effect transistor, which can generate pseudospin-polarized current. Also, we propose edge- and layer-pseudospin valves that operate analogously to conventional spin valves. The performance of valves in each pseudospin scheme is benchmarked by the pseudomagnetoresistance (PMR) ratio. The edge-pseudospin valve shows a nearly perfect PMR, with remarkable robustness against device parameters and disorder. These results may initiate new developments in pseudospin electronics.
\end{abstract}

%\pacs{}
\maketitle

\section{Introduction}
The internal degrees of freedom of electrons in nanostructures are an important focal point in modern condensed matter physics. In cases where these degrees of freedom are tunable by an external field, they can be employed in electronic devices for digital information processing~\cite{ney2003programmable,yuasa2004giant}. The most prominent example is electron spin. As spin couples to magnetic fields, it can be harnessed for spin electronics and quantum information applications~\cite{wolf2001spintronics,vzutic2004spintronics,awschalom2007challenges,
kane1998silicon,gershenfeld1997bulk}. Analogous applications can also be realized by exploiting other discrete degrees of freedom, referred to as the pseudospin. In materials such as multilayer graphene~\cite{novoselov2004electric} and transition-metal dichalcogenides (TMDs),~\cite{mak2010atomically} both the layer \cite{san2009pseudospin,min2008pseudospin,pesin2012spintronics,xu2014spin} and valley \cite{gunawan2006valley,rycerz2007valley,xiao2007valley,kim2014ultrafast,gong2013magnetoelectric} pseudospin degrees of freedom are present. In TMDs, the layer pseudospin is controlled by electrical polarization, while the spin and valley degrees of freedom can also be tuned by magnetic and optical means \cite{gong2013magnetoelectric}. Moreover, in TMDs, the strong coupling of layer and valley degrees of freedom offers a convenient platform for reliable spintronics implemented in two dimensions. Unfortunately, these features tend to disappear upon tailoring the material into a ribbon \cite{botello2009metallic}, which limits the system's possible applications in on-chip electronics. As a result, the quest for finding a practical, adjustable pseudospin degree of freedom in sub-two-dimensional nanostructures continues.

In this paper, we show that zigzag phosphorene nanoribbons (ZPNRs)  ~\cite{liu2014phosphorene,lu2014plasma} provide a platform for pseudospin electronics. In particular, charge transport in bilayer ZPNRs can be limited to one of the layers by applying a perpendicular electric field.  This gives rise to the concepts of ``up'' or ``down'' pseudospin when charge transport takes place in the top or bottom layer, respectively. Analogously, applying an in-plane electric field across either single layer or bilayer ZPNRs restricts  charge transport to the ``left'' or ``right'' edge of the ribbon, where electrons are considered to have ``left'' or ``right'' pseudospin polarization, respectively. In both cases, the pseudospin can be flipped by changing the sign of the applied electric field. In each case, we introduce nonmagnetic analogues of the spin field-effect transistor and spin valve, in which the role of the magnetic field is replaced by an electric field. For the pseudospin valves, a nonmagnetic counterpart of the magnetoresistance ratio, called the pseudomagnetoresistance ratio (PMR), is introduced and a large value is obtained at room temperature for both edge- and layer-pseudospin valve. Furthermore,  we evaluate how the PMR  is affected by the geometry of pseudospin-valves and presence of disorder. In particular, the PMR in the layer-pseudospin scheme is sensitive to the length of the valve and strong disorder, while in edge-pseudospin scheme is completely robust. These findings show that phosphorene nanoribbons provide key features necessary for pseudospin electronics that is compatible with current nanotechnology.

\section{Pseudospin Field-Effect Transistor}
\begin{figure}[ht]
  \begin{center}
  \includegraphics[width=.9\columnwidth]{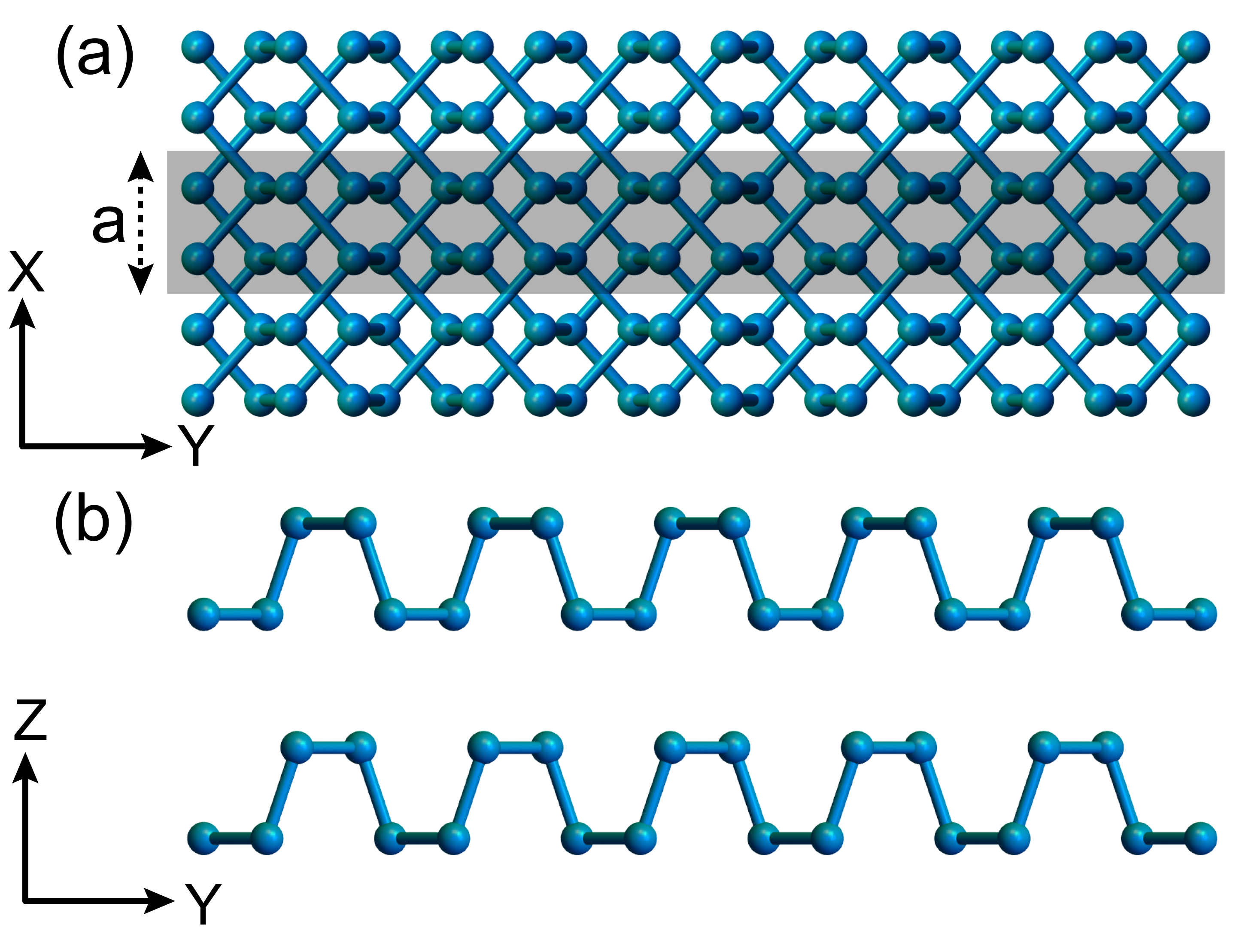}
  \end{center}
  \caption{\label{fig1} Schematic of a bilayer ZPNR. (a) Top view: The left and the right edges are zigzag. The gray rectangle denotes a unit cell for this ZPNR. (b) Side view: $a = 3.31$ $\mathrm{\AA}$ is the length of the unit cell.}
\end{figure}

\begin{figure}
  \begin{center}
   \includegraphics[width=.7\columnwidth]{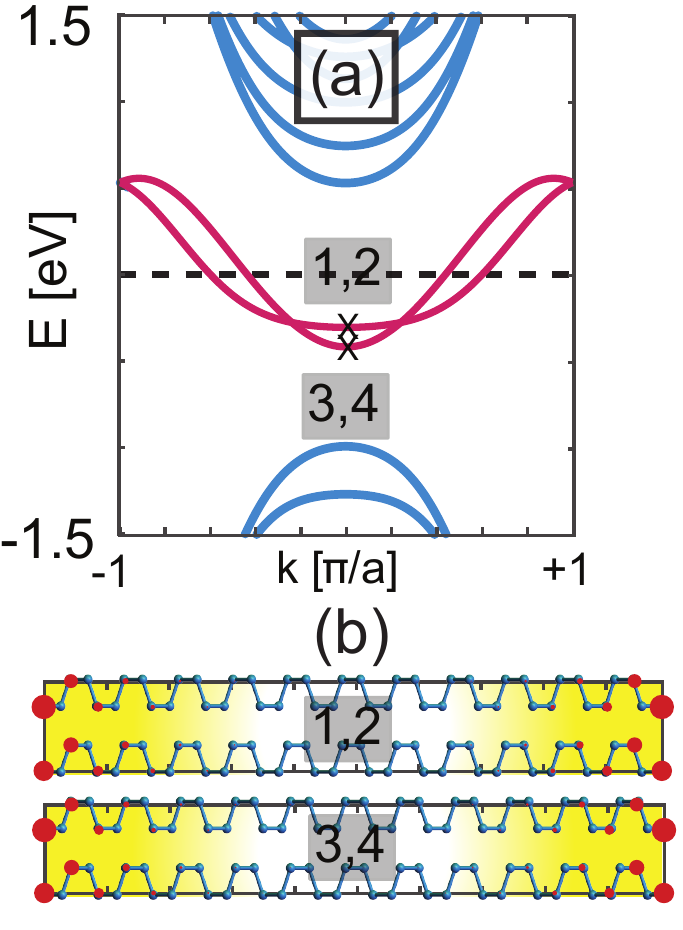}
  \end{center}
  \caption{\label{fig2} (a) Band structure of unbiased bilayer ZPNR. (b) The on-site probability density for the states with energies marked by ``x'' in panel (a). The red circles denote the probability density, with a larger circle diameter corresponding to higher probability density.}
\end{figure}

The crystal structure of a bilayer ZPNR is presented in Fig. \ref{fig1}. A unit cell is denoted by the solid gray rectangle. The length of the unit cell (a) is $3.31$ {\AA}. The band structure of a ZPNR, calculated using the fifteen-nearest-neighbors tight-binding Hamiltonian ~\citep{rudenko2015toward}, is shown in Fig. \ref{fig2}(a). The width of the ribbon is chosen to be $5$ nm. In this figure, one can see the presence of midgap bands (in red), disconnected from the bulklike bands (in blue). Each midgap band is twofold degenerate; therefore, a total of four midgap bands are present in the case of a bilayer ZPNR.  The Fermi level (dashed line) passes through the midgap bands, but is energetically far from the bulklike bands. Consequently, the ribbon is metallic, with low-field charge transport solely conducted through the states associated with the midgap bands. The probability density for the states associated with the four midgap bands at the zone center (wave number $k=0$), marked with $1-4$ in Fig. \ref{fig2}(a), is plotted across the ribbon in Fig. \ref{fig2}(b). The probability density peaks near the edges and decays toward the middle \cite{carvalho2014phosphorene}. The dispersion of the midgap bands and the localization of their corresponding wave function at the edges is not dependent on the width of the ribbon; rather, it is dictated by a topological invariant, fixed by the hopping elements of the Hamiltonian ~\cite{ezawa2014topological}.

%%%%%%%%%%%%%%%%%%%%%%%%%%%%%%%%%%%%%%%%%%%%
%  subsection?
%%%%%%%%%%%%%%%%%%%%%%%%%%%%%%%%%%%%%%%%%%%%
\begin{figure*}
  \begin{center}
   \includegraphics[width=1.9\columnwidth]{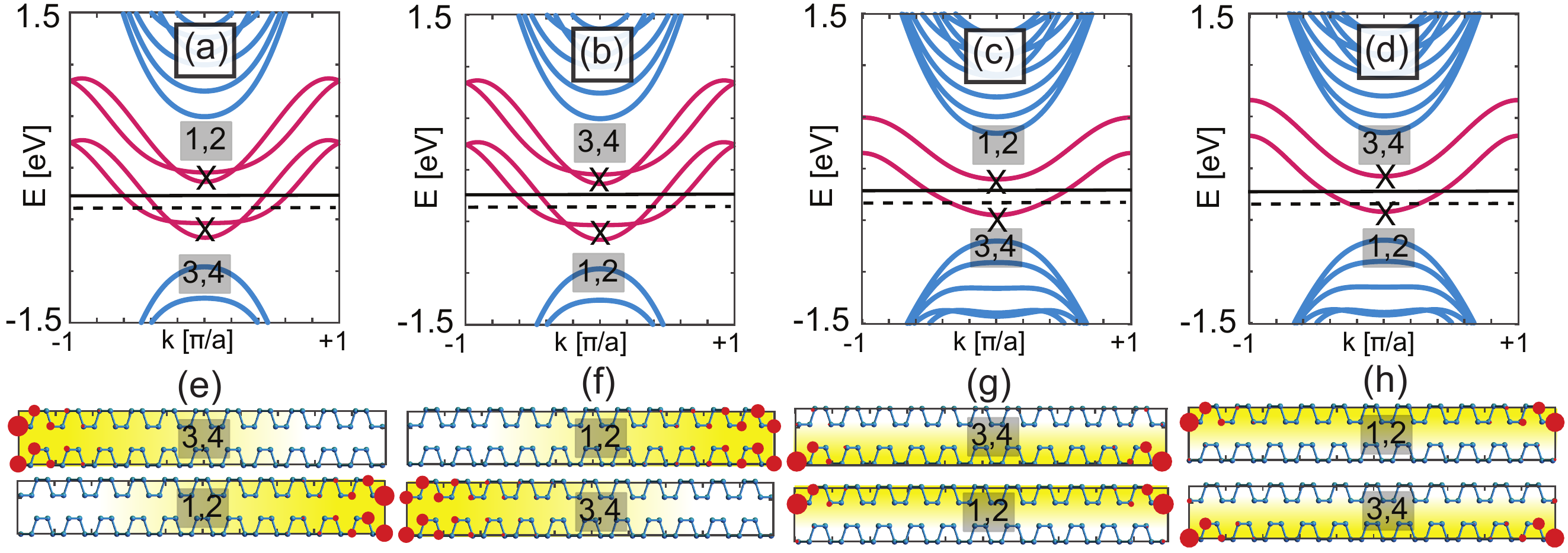}
  \end{center}
  \caption{\label{fig3} (a),(b) Band structure of bilayer ZPNR in the presence of an in-plane electric field in the width direction. In panel (a), the left edge of the ribbon is at $0.3$ V  and the right edge at $-0.3$ V. In panel (b), the signs of voltages have been flipped with respect to (a). (c),(d) Band structure of a bilayer ZPNR in the presence of a cross-plane (perpendicular) electric field.   In panel (c), the top of the ribbons is at $0.3$ V  and the bottom is biased to $-0.3$ V, while in (d) the signs of the voltages have been inverted. The probability density of the states marked in panels (a)--(d) are shown in the  corresponding panels of the second row, (e)--(h), respectively. The red circles denote the probability densities.  Larger circles correspond to higher probability densities.}
\end{figure*}
The position of the midgap bands in the energy diagram of ZPNRs can be shifted by applying an in-plane electric field (along the y-direction in Fig. \ref{fig1}, the ribbon width direction) or out-of-plane electric field (along z direction in Fig. \ref{fig1}). In both cases, the applied electric field can alter the energy of the midgap bands associated with the edge states. The effect of the electric field is incorporated in the tight-binding model through the diagonal terms of the Hamiltonian.  The ribbons are assumed to be infinitely long. The resulting band structure from applying an in-plane electric field in the width direction (y) is shown in Fig. \ref{fig3}(a). The right edge is fixed at the potential $+0.3$ V and the left fixed at $-0.3$ V and we assume that the potential varies linearly between the edges \cite{yuan2016quantum}. The applied bias moves bands 1 and 2 upward,  while it pushes bands 3 and 4 downward, and leaves the bulk bands unchanged. The probability densities of the states 1 and 2 are pushed to the left edge and those of states 3 and 4 are confined to the right edge [see Fig. \ref{fig3}(e)]. Under these circumstances, if the ribbon is connected to a source and drain, whose Fermi levels [one denoted by the dashed and the other by the solid horizontal line in Fig. \ref{fig3}(a)] are slightly offset with respect to one another owing to a small applied bias, the electronic transport can be channeled exclusively through the states 3 and 4, without contribution from any other states, so the current is carried only by the states associated with energy bands 3 and 4, which now have wave functions mostly located at the left edge of ZPNR [Fig. \ref{fig2}(e)]. Therefore, applying an in plane electric field along the width of the ZPNR leads to the generation of edge-pseudospin-polarized current in ZPNRs. In a similar manner, one can generate a current with the opposite pseudospin polarization by switching the sign of the applied bias on the two edges. When the right edge potential is fixed at $-0.3$ V instead of $+0.3$ V and the left edge bias is changed from $-0.3$ V to $+0.3$ V, states 3 and 4 move upward in the energy diagram, while 1 and 2 are pushed downward  [see Fig. \ref{fig3}(b)]. According to Fig. \ref{fig3}(b), at this bias, bands 1 and 2 are the only ones within the transport window designated by the source and drain Fermi levels and thus carrying current.  The corresponding states are at the right edge [Fig. \ref{fig3}(f)]. Therefore, a current with pseudospin right (carried only by electrons at the right edge) can be produced.

The feasibility of edge-pseudospin current generation is not limited to bilayer ZPNRs. The band-structure analysis given above can be expanded to single-layer ZPNRs, where there are two midgap bands instead of four. Applying an electric field along the width of a single-layer ZPNR leads to separation of these two midgap bands in energy. This separation in energy is associated with the confinement of the corresponding wave function at the opposite edges of the single-layer ZPNR. In the presence of a source and drain, where the transport window between the Fermi levels intersects only one midgap band, we can obtain an edge-pseudospin current, carried through the states localized near only one edge associated with that band.

\begin{figure}
  \begin{center}
   \includegraphics[width=.8\columnwidth]{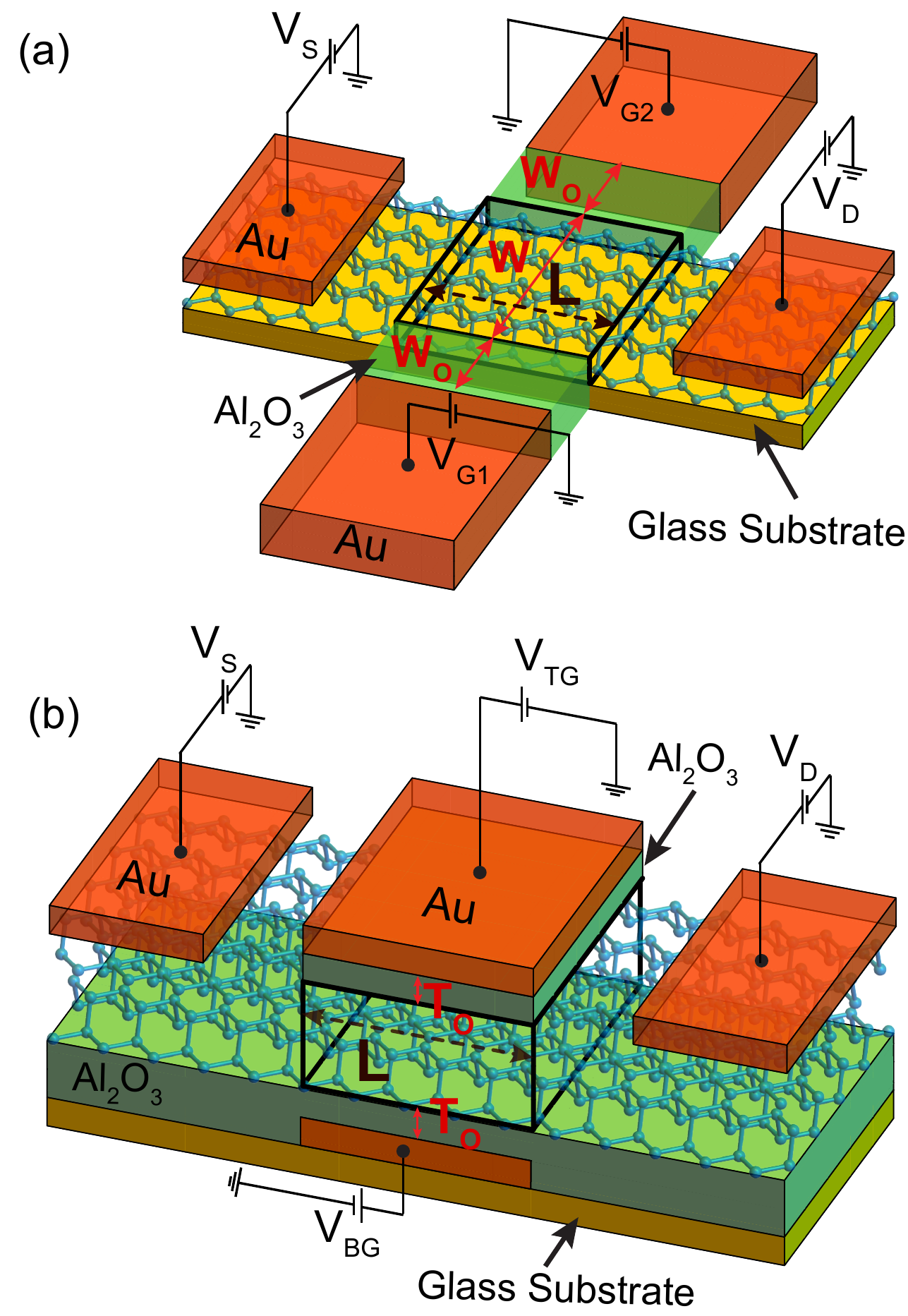}
  \end{center}
  \caption{\label{fig4} Schematic of the proposed field-effect transistor used to generate (a) edge-pseudospin and (b) layer-pseudospin current in ZPNRs. Gate oxides shown in green are $\mathrm{Al_{2}O_{3}}$ slabs. Gate electrodes, source, and drain are made of Au. The materials chosen are used in fabrication of bulk-phosphorene transistors and phosphorene dual-gate structures~\cite{tayari2016dual,kim2015dual}.}
\end{figure}

To conclude, in ZPNRs, edge-pseudospin-polarized current can be obtained by the application of a lateral (in-plane) electric field. This can realized by using a side-gate field-effect transistor (FET) [see Fig. \ref{fig4}(a)]. Here, a lateral field is created by applying voltages $V_{g1}$ and $V_{g2}$ to gates, which separates the states in the active region of the device (solid-line box in Fig. \ref{fig4}(a)). Also the carrier transport energy window is controlled by the biases on the source ($V_S$) and the drain ($V_D$). Side-gate FETs have been proven to be experimentally feasible as they have been used with graphene as the channel material for other applications~\cite{hahnlein2012side,molitor2007local}.

%%%%%%%%%%%%%%%%%%%%%%%%%%%
% subsection
%%%%%%%

If an applied electric field is perpendicular to the surface of the ribbon (along the z-direction in Fig. \ref{fig1}), similar midgap-band separation occurs, where the bands are pushed apart in pairs [see Fig. \ref{fig3}(c)]. Here, the voltage at the top of the ZPNR is fixed at $0.3$ V and at the bottom is $-0.3$ V and the voltages varies linearly in between. Once more, by attaching the ribbon to a source and drain, with the transport window between their Fermi levels chosen as shown in Fig. \ref{fig3}(c), current is carried solely by the states associated with bands 3 and 4.
Since the corresponding wave function are located in the bottom layer, the current generated can be referred to as ``pseudospin down'' current. Similar to the edge-pseudospin scheme, the pseudospin polarization of the generated current can be changed by switching the sign of the applied biases. With the top of the ribbon fixed at $-0.3$ V bias and the bottom at $+0.3$ V, bands 1 and 2 are in the carrier-transport energy window [see Fig. \ref{fig3}(d)];  their associated wave functions are located in the top layer [see Fig. \ref{fig3}(h)]. The current produced under these conditions will have pseudospin up (carried only by electrons at the upper layer).

The FET in Fig.\ref{fig4}(b) generates current with layer-pseudospin. Voltages on the top ($V_{TG}$) and bottom ($V_{BG}$) gates are applied to the active region of the device (solid line box).  This causes a potential difference (electric field) across the layers which leads to midgap band separation. The source and drain voltages tune the energy window of carrier transport. Dual-gate structures of this sort have been realized for bulk phosphorene transistors ~\cite{tayari2016dual,kim2015dual}. Layer-pseudospin current generation obviously cannot be achieved in single-layer ZPNRs.

\begin{figure}
  \begin{center}
   \includegraphics[width=.9\columnwidth]{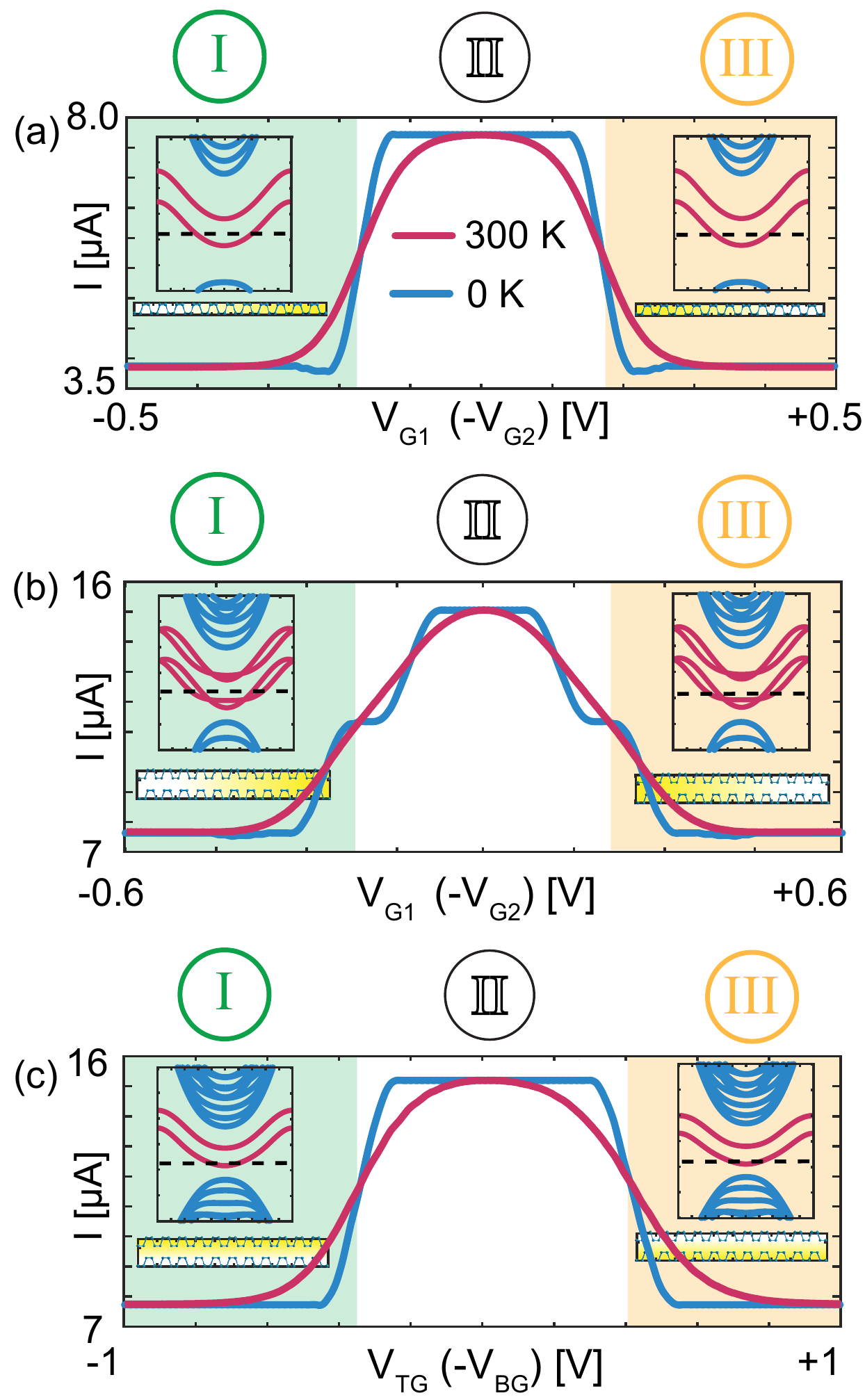}
  \end{center}
  \caption{\label{fig5} (a) and (b) I--V characteristics of the edge-pseudospin transistor portrayed in Fig.\ref{fig4}(a), where the channel material is the single-layer and bilayer ZPNR, respectively. (c) I--V characteristic of the layer-pseudospin transistor in Fig.\ref{fig4}(b).}
\end{figure}

To get a clearer understanding of the pseudospin FETs operation, we calculate electrical current using nonequilibrium Green's functions (NEGF) coupled with a Poisson solver~\cite{datta1997electronic}. Numerical implementation is described in the Appendix. All the simulations are done at room Temperature ($T = 300$ K).  Figures \ref{fig5}(a) and \ref{fig5}(b) show the current--voltage relation of the edge-pseudospin FET [schematic shown in Fig.\ref{fig4}(a)],  where single-layer and bilayer ZPNR were chosen as the channel material,  respectively. The Fermi level of source ($E_{fS}$) is set at $E_{fS} = -275$ meV and drain's Fermi level ($E_{fD}$) is kept at $E_{fD} = -325$ meV. $E_{fS,D}$ are offset with respect to one another by $50$ meV and have an average of $E = -0.3$ eV [this contact Fermi level is denoted by the dashed black line in the band structures portrayed in Fig. \ref{fig5}(a)-\ref{fig5}(c)]. $E_{fS,D}$  are kept at these values throughout the paper. In relation to Fig. \ref{fig4}(a), the ribbon width ($W$) and the gate-oxide widths ($W_o$) are $5$ nm and the ribbons are very long (we will discuss length effects further below). As can be seen, for both single and bilayer ZPNRs the edge-pseudospin field-effect device has three different regimes of operation, depending on the voltages on the side gates. At low bias [region II in Fig. \ref{fig5}(a) and (b)], the average contact Fermi level passes through all the the midgap states, leading to a current with no particular pseudospin polarization. By increasing the magnitude of gate voltages, the midgap bands are pushed apart. When the voltage on the gates is increased to highlighted region I (III), the narrow transport window close to the Fermi level passes through two out of the four midgap bands, whose wave function is confined to the right (left) edge. This results in a pseudospin-polarized current, with edge-pseudospin right (left).

The I--V curve of the layer-pseudospin FET [Fig. \ref{fig4}(b)] is shown in Fig. \ref{fig5}(c). For this structure, thickness of the gate oxide ($T_O$) is $5$ nm. The width of the bilayer ZPNR is $5$ nm and the active region is assumed very long. Similar to the case of edge-pseudospin FET, depending on the bias of the top gate $V_{TG}$ and bottom gate $V_{BG}$, the ribbon can be in three different regimes of operation. Once again, at low bias [region II in Fig.\ref{fig5}(c)] the current is carried by all the four midgap bands present in bilayer ZPNRs. This current does not have any particular pseudospin polarization. By tuning the bias voltages to region I (III),  the midgap bands  are separated in pairs, and only the midgap bands with their wave function in the upper (lower) layer overlap with the energy window imposed by source and drain Fermi levels. This results in a current with layer-pseudospin up (down).

In all cases discussed above, increasing the bias on the gates beyond regions I and III leads to further separation of midgap states. Consequently, the average  Fermi level of the contacts does not cross any bands, and the ribbon shows insulating behavior~\cite{soleimanikahnoj2016tunable}. These regions of operation are omitted from Figs. \ref{fig5}(a)-\ref{fig5}(c), as they occur at high electric field and bear no physical significance in the pseudospin scheme presented here. Simulations on wider ribbons ($W > 15$ nm) show that the regions of operations and the current--gate voltage curve remain the same as in Fig. \ref{fig5}, which underscores the topological nature of the midgap states that govern transport~\cite{ezawa2014topological}.

\begin{figure}
  \begin{center}
   \includegraphics[width=1.01\columnwidth]{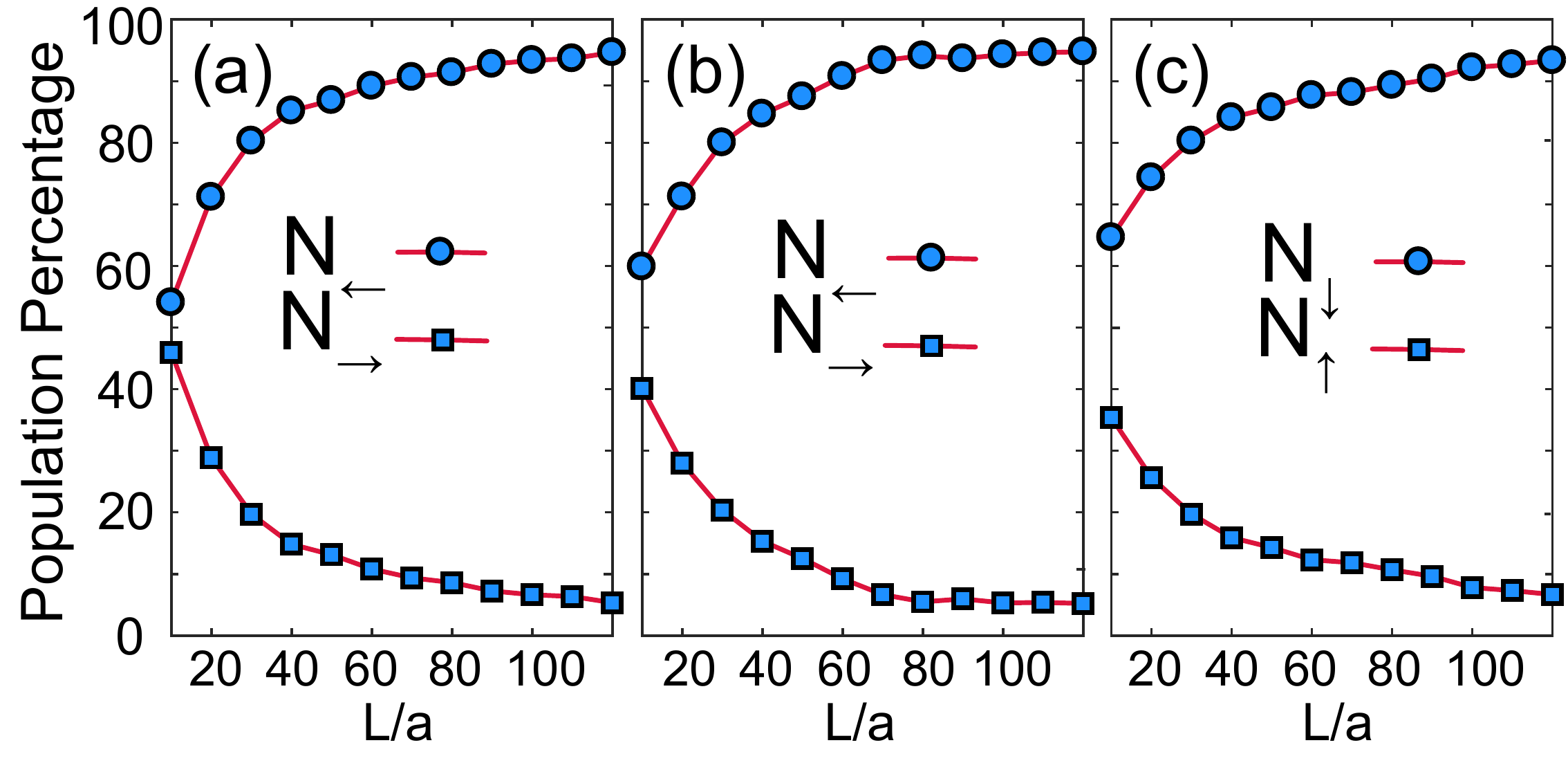}
  \end{center}
  \caption{\label{fig6} Population percentage of electrons in the edge- and layer-pseudospin field effect transistors shown in Figs. 4(a) and 4(b), respectively. Panels (a) and (b) correspond to edge-pseudospin FET where the single-layer and bilayer ZPNR were used, respectively. (c) Population percentage of electrons in the layer-pseudospin FET.}
\end{figure}

It is important to note that the response of the midgap bands to applied electric field depends on the length of the ribbon. Since the generation of the pseudospin current is based upon electric-field tuning of ZPNRs, the pseudospin polarization of the current is expected to be length dependent, as well. The polarization of the current can be measured by a population imbalance of electrons with opposite pseudospin in the active region of the pseudospin FET. Figure \ref{fig6} shows a population percentage of electrons as a function of length ($L$) in pseudospin FETs. Panels (a) and (b) correspond to edge-pseudospin FET with single layer and bilayer ZPNR as the channel material respectively. Voltages on the gates are tuned so that the FET generates a current with pseudospin-left ($V_{G1} = -V_{G2} = 0.35$ V). $N_{\rightarrow}$ ($N_{\leftarrow}$) is the population percentage of electrons in the right (left) edge. Figure \ref{fig6}(c) shows the percentage of electrons in layer-pseudospin FET in the top ($N_{\uparrow}$) and bottom ($N_{\downarrow}$) layer, where the FET gates are biased so pseudospin-down current is generated ($V_{TG} = -V_{BG} = 0.65$ V). As it is shown, with increasing the ribbon  length, electric field tuning of the midgap bands becomes more analogous to that in infinite-length ribbons and the population imbalance increases. In all cases,  length $L \geq 90a \simeq 29.7$ nm guarantees an electron population percentage of over $90$ \% ($N_{\leftarrow},N_{\downarrow} > 90\%$) of the expected pseudospin. Hence, the generation of current with high pseudospin polarization happens only when the quasi-one-dimensional nature of the ribbon is pronounced, i.e., when the ribbon is very long.

\section{Pseudospin Valve}
\begin{figure}
  \begin{center}
   \includegraphics[width=1\columnwidth]{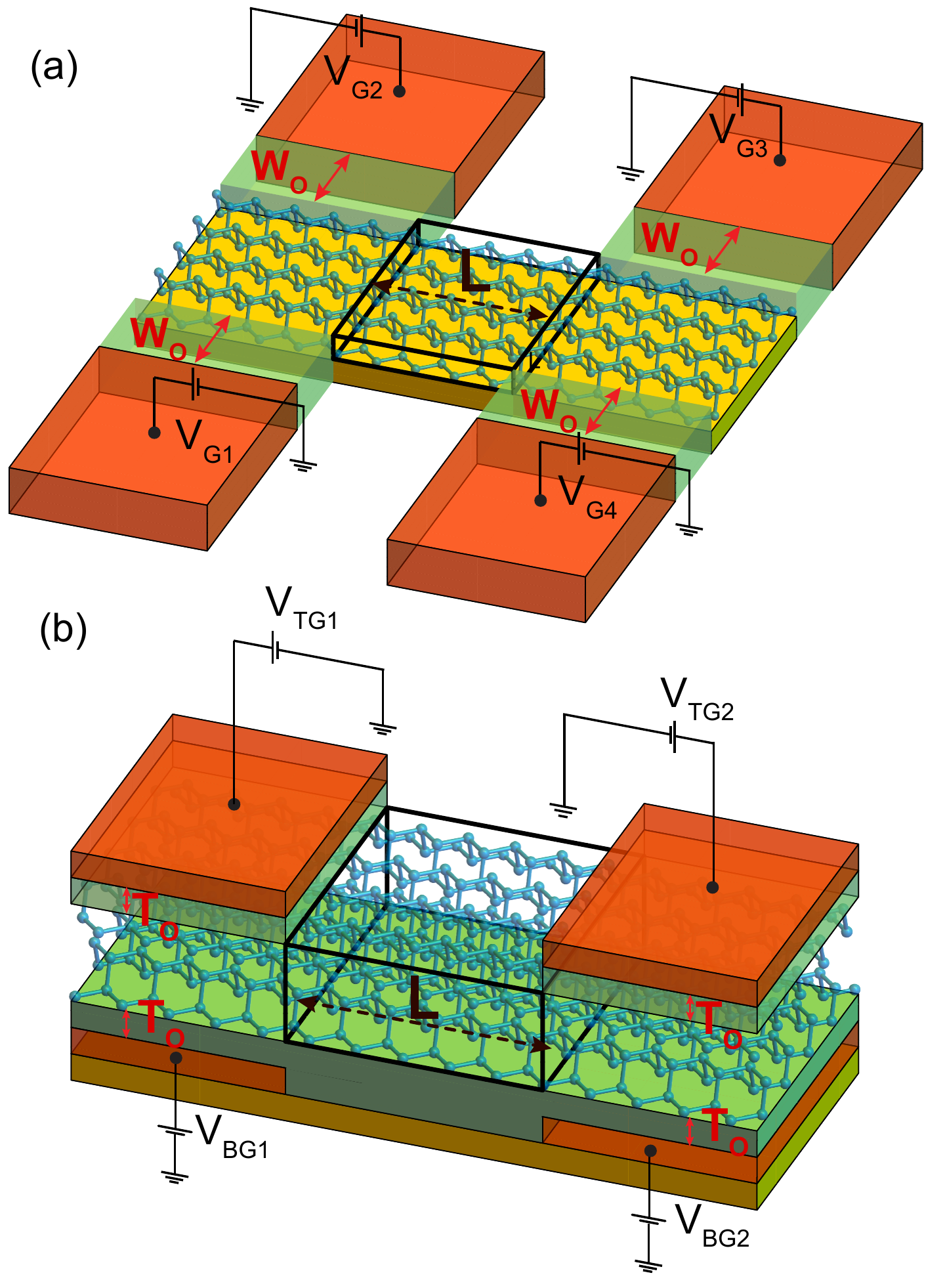}
  \end{center}
  \caption{\label{fig7} Schematics of (a) an edge-pseudospin valve and (b) a layer-pseudospin valve.}
\end{figure}

As in conventional spin devices, one would expect that the flow of the pseudospin-polarized current can be controlled by tuning the bias along the direction of electron transport ~\cite{baibich1988giant,viret1996spin}. This idea can be implemented through a pseudospin-based counterpart of the spin valve. Figure \ref{fig7}(a) shows the pseudospin valve in the edge scheme. The valve mode of operation changes through a variation of electrical resistance. The change in electrical resistance is controlled by the voltages on the gates at the opposite ends of the valve ($V_{G1-4}$). For instance, if in Fig.\ref{fig7}(a),  the voltages on gates 1 and 2 are opposites of each other ($V_{G1} = -V_{G2}$) and are tuned to region I (II) of Figs. \ref{fig5}(a),(b),   the ZPNR area sandwiched between gates 1 and 2 will carry current with edge-pseudospin left (right) only. If the second row of gates $\mathrm{G_{3,4}}$ are tuned to the same region of operation as the first row of gates, the electron pseudospin polarization will remain intact as they propagate through the valve. In this case the valve is said to be in the \textit{parallel configuration}. On the other hand, if the gates are tuned such that the region of operation changes as the current flows through the valve ($V_{G1,2}$ are tuned to region I and $V_{G3,4}$ biases are at region III or vice versa), the valve is in its \textit{antiparallel configuration}. In this configuration, by going through the valve, electrons are forced to ``rotate'' their pseudospin (going from the left to the right edge or vice versa). Considering that the overlap between the wave functions of electron at the opposite edges is vanishingly small, the forcible change of the edge-pseudospin in the antiparallel biasing configuration results in a much higher resistance than the resistance in the parallel configuration, where no edge switching of electrons is imposed.

A similar analysis can be given for the layer-pseudospin valve shown in Fig.\ref{fig7}(b). When all the
the gates are tuned according to Fig. \ref{fig5}(c) such that the regions at the opposite ends of the valve carry the same layer-pseudospin current, the device is said to be in its parallel configuration.

In the parallel configuration, electrons stay in the same layer while propagating through the valve intralayer electronic transport is dominant
and the conductance is high. However, if the gates at one end of the valve are biased to region I (II) and the the gates at the other end are tuned to region II (I), the valve is in the antiparallel configuration. In this case, electrons are forced to ``rotate'' their pseudospin, going from the upper (lower) layer to the lower (upper) layer. Considering that the intralayer hopping terms ($t_{\parallel}$) are much larger than their interlayer counterparts ($t_{\perp}$) ~\cite{rudenko2015toward}, electron interlayer movement in the antiparallel biasing configuration results in a much higher resistance than the intralayer movement characteristic of the parallel configuration. This resistance increase in the antiparallel configuration is analogous to the resistance increase due to spin scattering off domain walls in conventional spin devices in the antiparallel configuration~\cite{viret1996spin}.

In spintronic applications, the giant magnetoresistance (GMR) ratio is the standard for benchmarking spin-valve performance. Equivalently, the nonmagnetic version of the GMR ratio called the pseudomagnetoresistance (PMR) ratio characterizes the pseudospin-valve operation:
\begin{equation}\label{eq1}
  \mathrm{PMR} = \frac{R_{AP} - R_{p}}{R_{AP}}.
\end{equation}
Here, $R_{p}$ and $R_{AP}$ are the electrical resistance of the valve in parallel and antiparallel configurations respectively. For a perfect valve, the PMR ratio would be $100\%$.

Here, we calculated the PMR ratio for $5$-nm wide ZPNRs for both edge and layer pseudospin scheme using the self-consistent NEGF method described in the Appendix. The length of the active region ($L$) is $10a = 3.31$ nm.  The phosphorene areas sandwiched between the oxides are assumed to be infinitely long.  The oxide width ($W_O$) in the edge valve and the oxide thickness ($T_O$) in the layer valves are assumed to be 5 nm. All simulations are done at room temperature ($T = 300$ K).
\begin{figure}
  \begin{center}
   \includegraphics[width=.9\columnwidth]{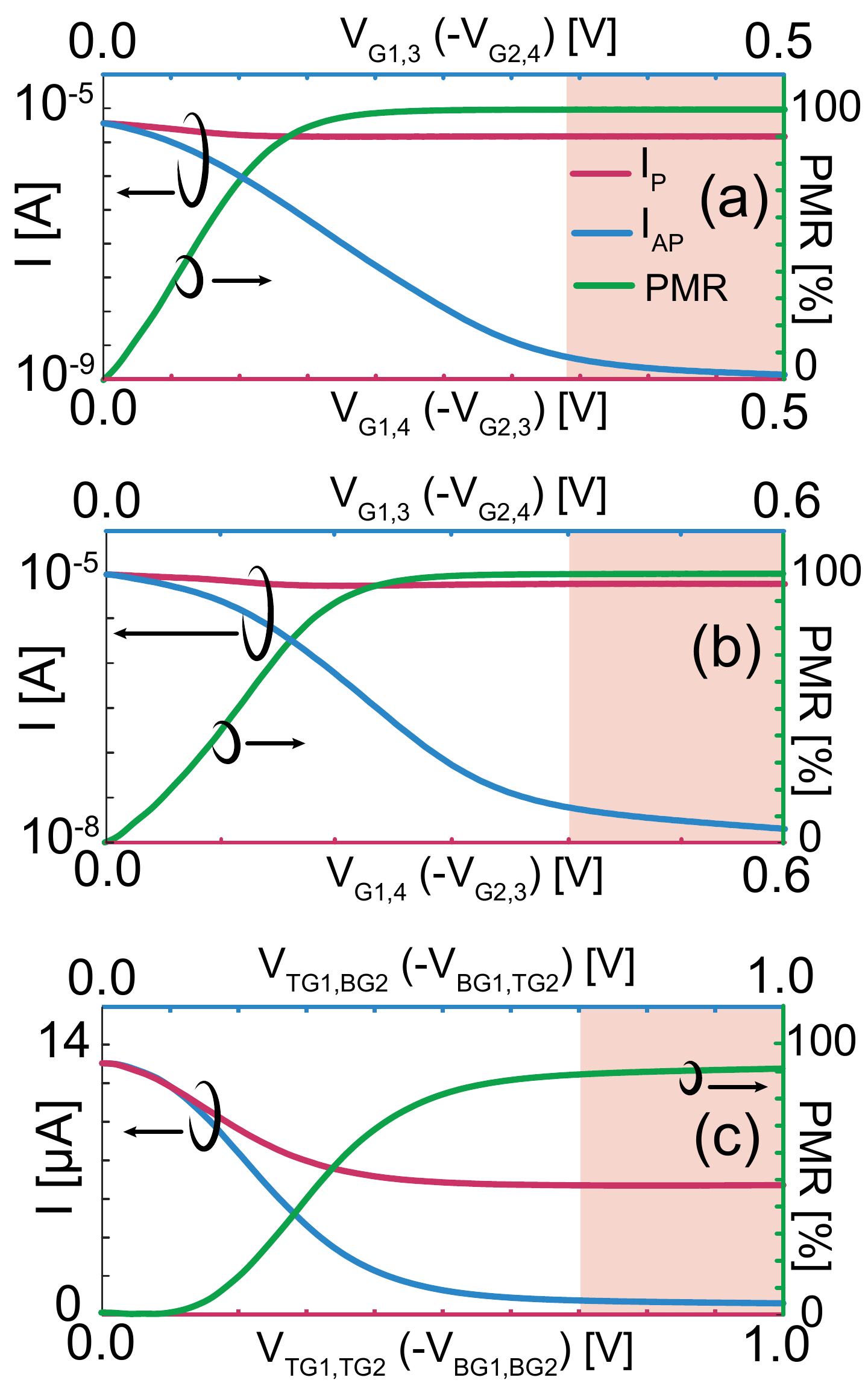}
  \end{center}
  \caption{\label{fig8} (a),(b) Current vs. gate voltage characteristic of (a) single-layer and (b) bilayer edge-pseudospin valve. (c) Current vs. gate voltage characteristic of the layer-pseudospin valve. In each panel, $I_P$ ($I_{AP}$) is the current in parallel (antiparallel) configuration. In all three panes, the pseudomagnetoresistance [\ref{eq1}] is shown by the green curve.}
\end{figure}

The calculated PMR ratio versus gate bias for the edge valve is shown in Figs. \ref{fig8}(a) and \ref{fig8}(b), where a single-layer ZPNR and a bilayer ZPNR were used as the channel material, respectively. The highlighted voltage-magnitude intervals in Fig. \ref{fig8}(a) is ($ 0.18$ V$ < |V_{G1-4}| < 0.50$ V) and in Fig. \ref{fig8}(b)  ($ 0.2$ V$ < |V_{G1-4}| < 0.60$ V)  are the regions of overlap between regions I and III from Figs. \ref{fig5}(a) and \ref{fig5}(b), respectively. Depending on the relative sign of applied biases,  if the ZPNR at both ends carries the same pseudospin current, the device is in its parallel configuration; the corresponding current ($I_P$) in this case is shown in Fig. \ref{fig8}. Likewise, if the region of operation changes along the valve, the device operates in its antiparallel configuration and a change in the pseudospin polarization is imposed; the corresponding current $I_{AP}$ is also shown in Fig. \ref{fig8}. As expected, $I_P$ is significantly higher than $I_{AP}$ in the highlighted voltage interval. This major difference is also mirrored in the PMR ratio, as shown in Figs. \ref{fig8}(a) and \ref{fig8}(b). As can be seen, for both single-layer and bilayer ZPNR in the highlighted region, the PMR ratio exceeds $99\%$, indicating a nearly perfect valve. Similarly, the PMR ratio versus gate bias for the layer valve is portrayed in Fig. \ref{fig8}(c), where the highlighted region ($ 0.4$ V$ < |V_{TG1,2}|,|V_{BG1,2}| < 1.0$ V) is the overlap between region I and III in Fig. \ref{fig5}(c). Inside the overlap region, when both ends of the valve are in the same region of operation the valve is in parallel configuration. Conversely, if the region of operation changes along the charge transport direction the valve is in antiparallel configuration. The imposed change of layer-psudospin in antiparallel configuration leads to a considerable difference between the current in this case compared to parallel configuration [see Fig. \ref{fig8}(c)]. The difference between $I_P$ and $I_{AP}$ leads to a large value of the PMR ratio ($>92\%$) inside the highlighted region shown in Fig.\ref{fig8}(c).

\begin{figure}
  \begin{center}
   \includegraphics[width=1\columnwidth]{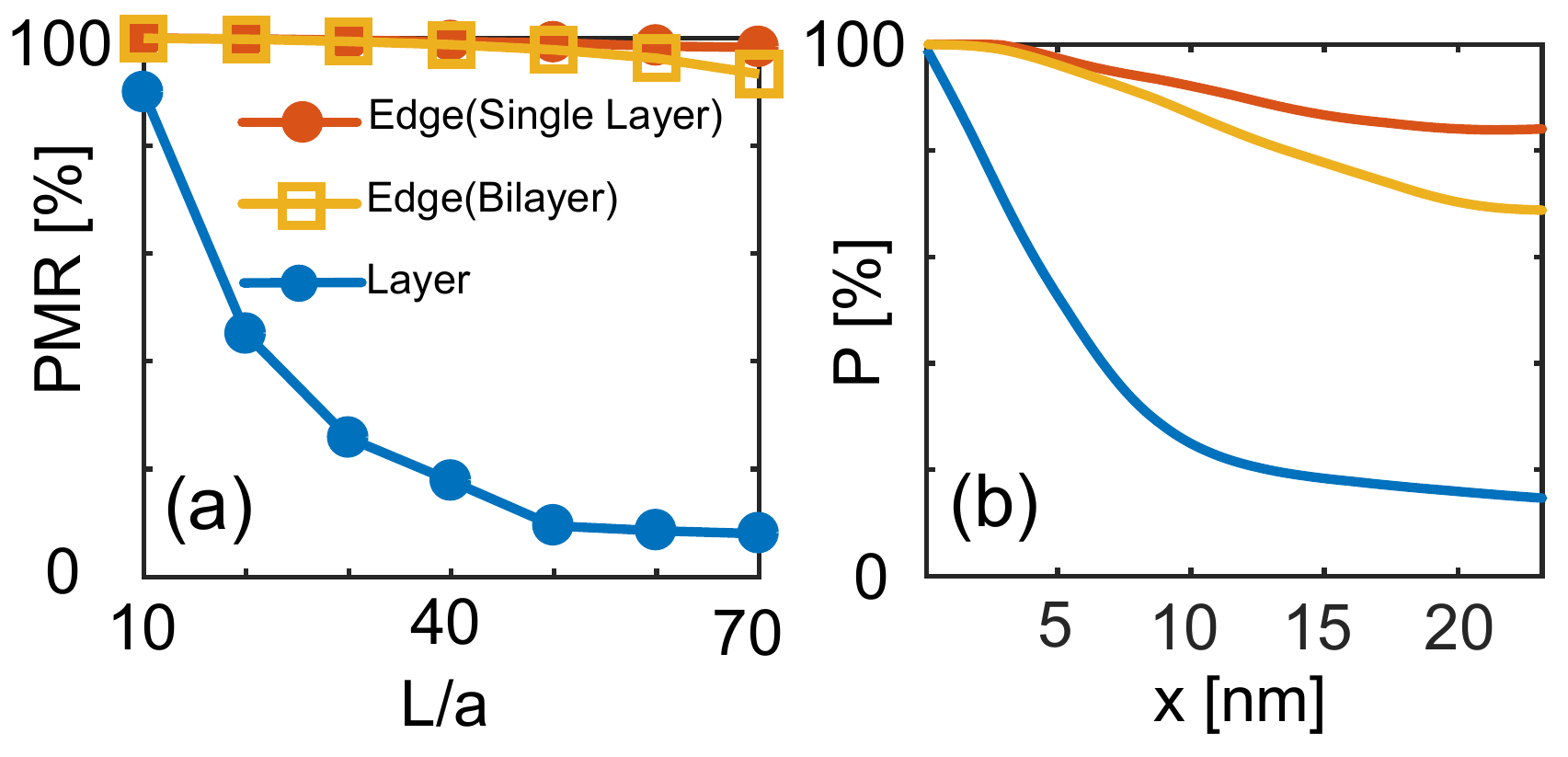}
  \end{center}
  \caption{\label{fig9} (a) The pseudomagnetoresistance ratio and (b) polarization as a function of valve length in the edge and layer schemes. For the edge-pseudospin valve $|V_{G1-4}| = 0.35$ V for both single-layer and bilayer ZPNR. For the layer-pseudospin valve $|V_{TG1,2}| = |V_{BG1,2}| = 0.65$ V.}
\end{figure}

In conventional spin valves, the magnetoresistance ratio tends to decrease by increasing the length of the valve~\cite{shim2008large,barraud2010unravelling}. The PMR ratio, being the nonmagnetic version of magnetoresistance ratio, also decreases as the length of the valve increases. Figure \ref{fig9}(a) shows the PMR ratio as a function of the valve length ($L$) in the layer and edge schemes. The PMR ratio of the layer valve drops rapidly as a function of length, demonstrating the short relaxation length of the layer-pseudospin degree of freedom. In contrast, the edge-pseudospin valve maintains its high value of the PMR ratio even for large lengths ($L = 70a \simeq 23.2$ nm) for both single-layer and bilayer ZPNRs,  showing the long relaxation length of the edge-pseudospin degree of freedom. This originates from the small overlap of the electron wave functions at the opposite edges of the ZPNRs, which makes transport of electrons from one edge to the other highly improbable. As a result, electrons tend to keep their edge-pseudospin polarization over much longer distances than compared to their layer-pseudospin counterpart.

The relaxation of the pseudospin along the valve can also be traced to the population imbalance between electrons with opposite pseudospins, as they propagate through the valve. The population imbalance of electrons with opposite edge-pseudospin can be represented using the pseudospin polarization, $P(x)$, defined as:

\begin{equation}
P(x) = \frac{|n_{\leftarrow}(x) - n_{\rightarrow}(x)|}{n_{\leftarrow}(x) + n_{\rightarrow}(x)}.
\end{equation}

\noindent
Here, $n_{\leftarrow}(x)$ ($n_{\rightarrow}(x)$) is the population of electrons with edge-pseudospin left (right) at position $x$ along the valve. The polarization of layer-pseudospin is the same as above with $n_{\substack{\leftarrow \\ \rightarrow}}$ replaced with  $n_{\substack{\uparrow  \downarrow}}$. Figure \ref{fig9}(b) shows the pseudospin polarization of electrons along the length for edge- and layer-pseudospin valve. Here, $L = 70a \simeq 23.1$ nm. As can be seen, electrons are injected from the left with a nearly perfect pseudospin polarization ($P \simeq 100 \%$). As they move through the valve, the population imbalance of electrons with opposite pseudospin decreases. As a result, pseudospin polarization of electrons decreases. The drop in the polarization of layer-pseudospin is more significant than for edge-pseudospin. This verifies the fast relaxation of layer-pseudospin in comparison with edge-pseudospin.

As the layer-valve works based on the asymmetry between the interlayer and intralayer hopping, it is also realizable in other bilayer Van der Waals materials. In such valves, the PMR ratio and polarization of pseudospin would be dependent on the value of the interlayer hopping elements ($t_{\perp}$) with respect to intralyer hopping elements ($t_{\parallel}$). Assuming the hopping elements of phosphorene,  preliminary calculation on a valve with $L = 70a$ shows that reducing the interlayer hopping elements by 33.3\% ($t_{\perp} \rightarrow \frac{2}{3} t_{\perp}$) improves the PMR ratio from 8\% to 34\% and the polarization at the end of the valve from 16\% to 41\%. This shows that layered materials with a weaker Van der Waals force between the layers are better candidates for layer-pseudospin electronics.

\begin{figure}
  \begin{center}
   \includegraphics[width=.8\columnwidth]{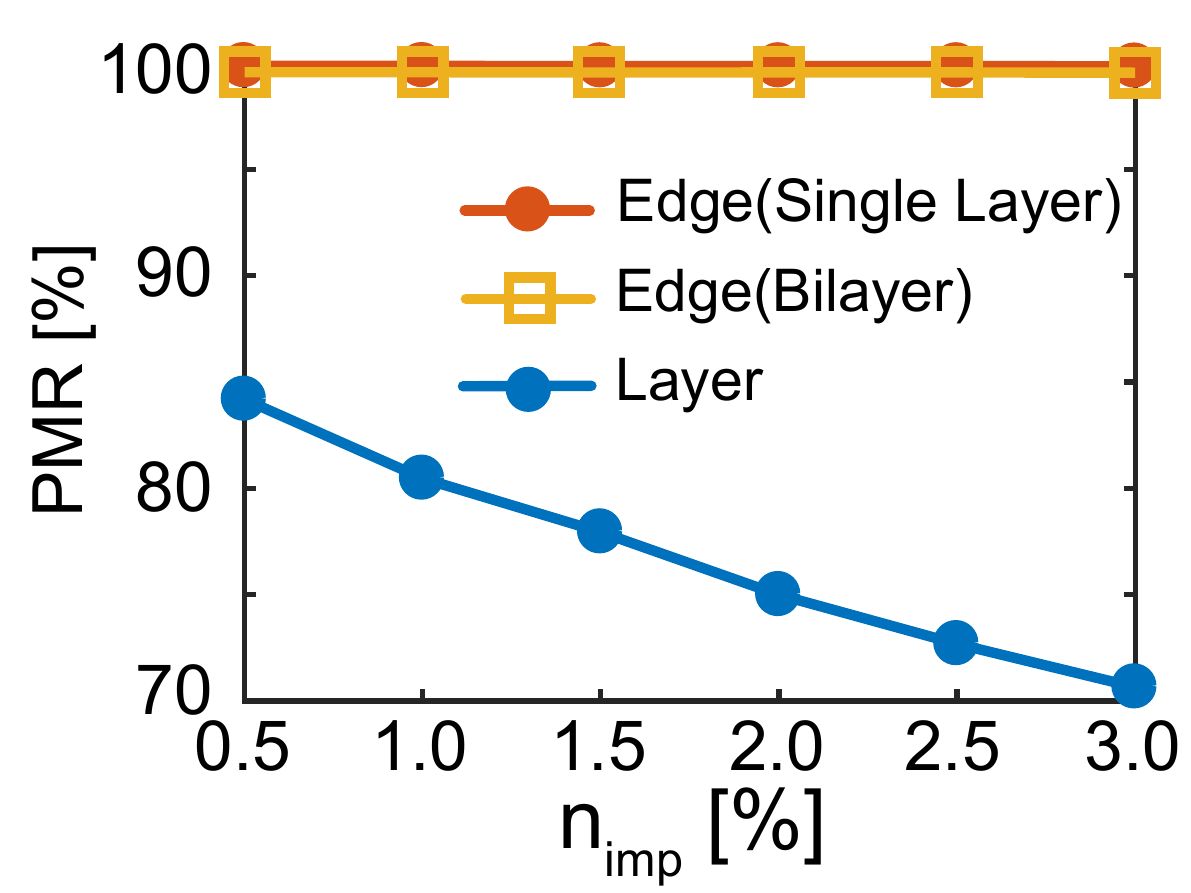}
  \end{center}
  \caption{\label{fig10} Pseudomagnetoresistance ratio as a function of impurity abundance. Edge-pseudospin valve biases are $|V_{G1-4}| = 0.35$ V for both single-layer and bilayer ZPNR. For the layer-pseudospin valve,  $|V_{TG1,2}| = |V_{BG1,2}| = 0.65$ V. Data points are
from numerical simulation with $\delta U = 1$ eV and $\xi = 5$ \AA .}
\end{figure}

As a further matter, phosphorene samples are found to be sensitive to the environment, which makes the role of impurities significant~\cite{wood2014effective,island2015environmental}. In particular, potential fluctuations caused by charged impurities play a crucial role in electronic transport of two-dimensional materials~\cite{ong2014anisotropic,dean2010boron,ni2009probing}. Here, the effect of charged impuirities is added to our model in the form of superposition of Gaussian potential fluctuations~\cite{paez2016disorder},
\begin{equation}\label{eq2}
U(\mathbf{r}_i) = \sum_{k=1}^{N_{\mathrm{imp}}}U_ke^{|\mathbf{r}_i - \mathbf{R}_k|^2/2\xi^2}.
\end{equation}
Here, $\mathbf{r}_i$ denotes the position of the lattice site $i$. $N_{\mathrm{imp}}$ is the number of scatterers that are located at $\{\mathbf{R}_k\}_{k = 1,N_{\mathrm{imp}}}$. These locations are chosen by a random uniform distribution. The scatterers have amplitudes $\{U_k\}_{k = 1,N_{\mathrm{imp}}}$ taken from a uniform distribution $[-\delta U/2,\delta U/2]$. $\xi$  is the correlation length. Density of scatterers is $n_{\mathrm{imp}} = N_{\mathrm{imp}}/N$ where $N$ is the number of phosphorus atoms in the active region of the valve. At each point of the lattice, the potential term from Eq. (\ref{eq2}) is calculated and added to the diagonal term of the main Hamiltonian that corresponds to that lattice point. This model has been previously used for modeling of charged impurities in phosphorene \cite{paez2016disorder} and graphene~\cite{mucciolo2010disorder,mivskovic2012ionic,wehling2009impurities,
vierimaa2017scattering}, where a good agreement with experiment was obsereved \cite{tan2007measurement,adam2008density}. Figure \ref{fig10} shows the PMR ratio as a function of impurity density ($n_{\mathrm{imp}}$). $\delta U = 1$ eV and $\xi$ is fixed at 5 \AA . The values within this range were shown to capture the role of charged-impurity in carrier transport of nanostrips effectively~\cite{mucciolo2010disorder}. Each data point is an average over 200 configurations. The PMR ratio in the layer-pseudospin scheme drops as the number of scatterers increases.  In contrast, the PMR ratio for the edge valve is completely robust against $n_{\mathrm{imp}}$.
The reason for this difference between the layer and edge valves originates from the mechanisms based on which they operate. The layer valve works based on the imbalance between interlayer and intralayer electron transport in ZPNRs. By increasing the scatterer density, the difference between interlayer and intralayer hopping elements is overshadowed by the scattering in a larger area of the ribbon. As a result, the layer PMR ratio decreases. In contrast, the edge-valve operation is based upon the very existence of edge states, which endure in the presence of scatterers ~\cite{ryu2002topological}. In particular, the existence of edge states in ZPNRs is associated with a topological winding number, which is independent of the diagonal elements of the Hamiltonian. As the presence of disorder changes only the diagonal terms of the Hamiltonian, edge states remain intact. Hence, the edge PMR ratio is unchanged in the presence of impurities.

In general, implementation of conventional spintronic applications requires  efficient generation and detection of spin-polarized current. The former is typically obtained by generating a spin current in a magnetically tuned ferromagnetic nanofilm, followed by injection into a semiconductor via an ohmic contact; the latter is carried out by the spin valve. The pseudospin schemes described in this paper integrate the generation and detection of pseudospin current into a single material, which makes for easier miniaturization. The applications of spintronics, such as spin-logic gates, are found to be attainable in pseudospin schemes~\cite{schaibley2016valleytronics}. Also, the advantages of spin-based logic devices over conventional metal-oxide semiconductor (MOS) field-effect devices in terms of power consumption and speed are also expected in their pseudospin-based counterparts~\cite{banerjee2009bilayer}.

The pseudospin devices discussed in this paper can also be realized using the recently discovered skewed-armchair phosphorene nonoribbons~\cite{soleimanikahnoj2016tunable}. Similar to ZPNRs, midgap states are present in the energy dispersion of skewed-armchair phosphorene nanoribbons, which will facilitate pseudospin electronics.

\section{Conclusions}

In summary, electron transport in metallic ZPNRs is governed by the states localized near the edges, whose energies belong to the midgap bands that are energetically far from the bulk bands. These states can be electrically manipulated by gating ZPNRs in two different ways, which bring about two practical versions of the pseudospin. One is the edge pseudospin, where pseudospin ``left'' (``right'') is associated with the conducting electrons located near the left (right) edge. The other is realized in bilayer ZPNRs,  where limiting electron transport to the ``top'' (``bottom'') layer gives rise to the concept of ``up'' (``down'') pseudospin. In each scheme, we proposed two devices: an FET for the generation of pseudospin-polarized current and a pseudospin valve. The PMR ratio is calculated for both edge and layer-pseudospin valves, where the edge-pseudospin valve is nearly perfect and robust against variations in device parameters and disorder. Although the results presented here are promising, we acknowledge the experimental challenges in the fabrication of zigzag phosphorene nanoribbons with perfect edges. Nevertheless, nanoribbons with atomically precise edges have been realized in graphene~\cite{cai2010atomically} and TMDs~\cite{chen2017atomically} which belong to the same family of two-dimensional honeycomb-lattice materials as phosphorene.
Thus, the results presented here should be viewed as a start, which will hopefully encourage further experimental and theoretical work.

\section*{Acknowledgments}
The authors gratefully acknowledge support by the US Department of Energy under Award No. DE-SC0008712 (Office of Basic Energy Sciences, Division of Materials Sciences and Engineering, Physical Behavior of Materials Program). Related preliminary work was funded as a seed project by the UW--Madison MRSEC (NSF award DMR-1121288). The work was performed using the compute resources and assistance of the UW-Madison Center for High Throughput Computing (CHTC) in the Department of Computer Sciences. The authors thank Farhad Karimi for his valuable comments.

\section{Appendix:
IMPLEMENTATION DETAILS}
The modeling of electrical devices was performed through a self-consistent solution of two equations. The first is the retarded Green's $G$ function, which describes the dynamics of electrons inside the active region of the devices~\cite{datta1997electronic}
\begin{equation}
G_{r,r'}(E) = [E - H_{r,r'} - U_{r,r} - \Sigma^S_{r,r'}(E)  - \Sigma^D_{r,r'}(E)].
\end{equation}
$E$ is the energy at which the Green's function is being calculated. $r$ and $r'$ are the lattice-point positions. $H$ is the Hamiltonian of the active region and $U$ is the self-consistent potential being applied to the active region. $\Sigma^{S(D)}$ is the self energy of the source (drain). The effects of metal contacts have been ignored and the contact self-energies were calculated using the Sancho-Rubio iterative scheme~\cite{sancho1985highly}. The number of electrons ($n$) and holes ($p$) at each lattice point is calculated using $ n(r)= 2\int \frac{dE}{2\pi}G^n_{r,r}$ and $h(r) = 2\int \frac{dE}{2\pi}G^p_{r,r}$, respectively. Here, $G^n_{r,r'}$ and $G^p_{r,r'}$ are analytical functions: $\Gamma^{S(D)} = -2Im[\Sigma^{S(D)}]$, the Green's function, the Fermi level of the source ($E_{fS}$) and the drain ($E_{fD}$), and the Fermi-Dirac distribution function ($f(E)$):
\begin{subequations}
\begin{equation}
G^n =  G[\Gamma^Sf(E-E_{fS}) + \Gamma^Df(E-E_{fD})]G^{\dagger},
  \end{equation}
\begin{equation}
G^p =  G[\Gamma^S(1 - f(E-E_{fS})) + \Gamma^D(1 - f(E-E_{fD}))]G^{\dagger}.
  \end{equation}
\end{subequations}

The second equation that was solved self-consistently with the first is Poisson's equation:
\begin{equation}
\nabla(\epsilon(r)\nabla U(r)) = -\rho (r)
\end{equation}
 which determines the self-consistent potential $U$ for a given charge distribution $\rho (r)$. Charge distribution is a function of the electron and hole occupation obtained from the Green's function. The dielectric function  was assumed to be position dependent to take into account the different materials (phosphorene sheet; $\mathrm{Al_2O_3}$). In the simulations, the dielectric function of $\mathrm{Al_2O_3}$ was assumed to be $9.5\epsilon_0$ and those of single-layer and bilayer phosphorene were taken to be $1.12\epsilon_0$ and $1.72\epsilon_0$ respectively~\cite{wang2015native}. The effect of substrate was ignored and gates were taken into account via  Dirichlet boundary condition. Neumann boundary conditions were assumed at the remaining boundaries. 
 
 The Green's function and the Poisson equation were solved self-consistently until convergence was obtained. The converged Green’s function was then used to obtain the current ($I$):
\begin{equation}
I = \frac{2q}{h}\int_{-\infty}^{+\infty}dE \ T\left(E\right)[f(E-E_{fS}) - f(E-E_{fD})]. 
\end{equation}
Here, $T(E)$ is the transmission function, calculates as 
\begin{equation}
    T = Trace(\Gamma^SG\Gamma^DG^{\dagger}).
\end{equation}

%\bibliography{RefCopy}

%

 % input acknowledgement

%\bibliography{Ref}
\end{document}